\documentclass[aps,twocolumn]{revtex4}

\begin{document}
\title{On the vector solutions of Maxwell equations \\
in spherical coordinate systems}
\author{E.A. Matute}
\affiliation{Departamento de F\'{\i}sica, Universidad de Santiago de
Chile, \\ Casilla 307 - Correo 2, Santiago, Chile, \\
e-mail: ematute@lauca.usach.cl}
\begin{abstract}
\noindent The Maxwell equations for the spherical components of
the electromagnetic fields outside sources do not separate into
equations for each component alone.   We show, however, that
general solutions can be obtained by separation of variables in
the case of azimuthal symmetry.  Boundary conditions are easier to
apply to these solutions, and their forms highlight the
similarities and differences between the electric and magnetic
cases in both time-independent and time-dependent situations.
Instructive examples of direct calculation of electric and
magnetic fields from localized charge and current distributions
are presented.
\\[7pt]
\\[10pt]
\noindent Las ecuaciones de Maxwell para las componentes
esf\'ericas de campos electromagn\'eticos en regiones libres de
fuentes no son separables en ecuaciones para cada una de sus
componentes.  Se muestra, sin embargo, que soluciones generales
pueden ser obtenidas por separaci\'on de variables en el caso de
simetr\'{\i}a azimutal.  Las condiciones de borde son f\'aciles de
aplicar para estas soluciones, y sus formas destacan las
similitudes y diferencias entre los casos el\'ectrico y
magn\'etico, tanto para las situaciones independientes del tiempo
como para las de tiempo dependientes.  Se presentan ejemplos
instructivos de c\'alculos directos de campos el\'ectricos y
magn\'eticos producidos por distribuciones localizadas de cargas y
corrientes.
\\[7pt]
\emph{Descriptores:} Ecuaciones de Maxwell; coordenadas esf\'ericas;
campos el\'ectrico y magn\'etico; problemas con condiciones de borde.
\\[10pt]
\end{abstract}
\maketitle

\renewcommand{\thesection}{\arabic{section}}

\section{Introduction}
The Maxwell equations for the electromagnetic field vectors, expressed in
the International System of Units (SI), are~\cite{Panofsky}
\begin{eqnarray}
& & {\bf \nabla} \cdot {\bf D} = \rho , \hspace{1cm} {\bf \nabla}
{\bf \times} {\bf E} = - \displaystyle \frac{\partial {\bf
B}}{\partial t} , \nonumber \\ & & \nonumber \\ & & {\bf \nabla}
\cdot {\bf B} = 0 , \hspace{1cm} {\bf \nabla} {\bf \times} {\bf H}
= {\bf J} + \displaystyle \frac{\partial {\bf D}}{\partial t} ,
\label{Maxwell}
\end{eqnarray}
where the source terms $\rho$ and ${\bf J}$ describe the densities of
electric charge and current, respectively.  For a linear, isotropic
medium ${\bf D}$ and ${\bf H}$ are connected with the basic fields
${\bf E}$ and ${\bf B}$ by the constitutive relations
\begin{equation}
{\bf D} = \epsilon {\bf E} , \;\;\;\; {\bf H} = {\bf B}/\mu ,
\label{linear}
\end{equation}
where $\epsilon$ and $\mu$ are the permittivity and permeability of the
medium, respectively.

The boundary conditions for fields at a boundary surface between two
different media are~\cite{Panofsky+}
\begin{eqnarray}
& & {\bf n} \cdot ({\bf D}_{1} - {\bf D}_{2}) = \rho_{S} ,
\hspace{0.5cm} {\bf n} {\bf \times} ({\bf E}_{1} - {\bf E}_{2}) =
{\bf 0} , \nonumber
\\ & & \nonumber \\
& & {\bf n} \cdot ({\bf B}_{1} - {\bf B}_{2}) = 0 , \hspace{0.7cm}
{\bf n} {\bf \times} ({\bf H}_{1} - {\bf H}_{2}) = {\bf J}_{S} ,
\label{boundary}
\end{eqnarray}
where $\rho_{S}$ and ${\bf J}_{S}$ denote the surface charge and
current densities, respectively, and the normal unit vector ${\bf n}$
is drawn from the second into the first region.  The interior and
exterior fields satisfy the homogeneous vector wave equations
\begin{eqnarray}
& \nabla^{2} {\bf E} - \epsilon \mu \; \displaystyle
\frac{\partial^2 {\bf E}}{\partial t^2} = {\bf 0} , & \nonumber \\
& & \nonumber \\ & \nabla^{2} {\bf B} - \epsilon \mu \;
\displaystyle \frac{\partial^2 {\bf B}}{\partial t^2} = {\bf 0} ,
& \label{Dalembert}
\end{eqnarray}
which are obtained from Eqs.~(\ref{Maxwell}) and (\ref{linear})
for regions free of charge and current by combining the two curl
equations and making use of the divergence equations together with
the vector identity
\begin{equation}
\nabla^{2} (\;\;) = {\bf \nabla} ({\bf \nabla} \cdot \;\;) -
{\bf \nabla} {\bf \times} ({\bf \nabla} {\bf \times} \;\;) .
\label{nablas}
\end{equation}
Changes in the electromagnetic fields propagate with speed
$v=1/\sqrt{\epsilon \mu}$ .

Without any loss of generality, we may consider only harmonic time
dependence for sources and fields:
\begin{eqnarray}
& \rho({\bf r},t) = \rho({\bf r}) e^{-i \omega t} , \hspace{0.5cm}
{\bf J}({\bf r},t) = {\bf J}({\bf r}) e^{-i \omega t} , & \nonumber \\
& & \nonumber \\ & {\bf E}({\bf r},t) = {\bf E}({\bf r}) e^{-i
\omega t} , \hspace{0.5cm} {\bf B}({\bf r},t) = {\bf B}({\bf r})
e^{-i \omega t} , & \label{harmonic}
\end{eqnarray}
where the real part of each expression is implied.  Equation
(\ref{Dalembert}) then becomes time-independent:
\begin{equation}
\nabla^{2} {\bf E} + k^2 \; {\bf E} = 0 ,  \hspace{0.5cm}
\nabla^{2} {\bf B} + k^2 \; {\bf B} = 0 , \label{Helmholtz}
\end{equation}
where $k^2=\epsilon\mu\omega^2$.  These are vector Helmholtz
equations for transverse fields having zero divergence.  Their
solutions subject to arbitrary boundary conditions are considered
more complicated than those of the corresponding scalar equations,
since only in Cartesian coordinates the Laplacian of a vector
field is the vector sum of the Laplacian of its separated
components.  For spherical coordinates, as for any other
curvilinear coordinate system, we are faced with a highly
complicated set of three simultaneous equations, each equation
involving all three components of the vector field.  This
complication is well known and general techniques for solving
these equations have been developed, based on a dyadic Green's
function which transforms the boundary conditions and source
densities into the vector solution~\cite{Morse}. We shall show,
however, that in the case of spherical boundary conditions with
azimuthal symmetry, the solution can be obtained more conveniently
by means of separation of variables. Several applications of
physical interest can then be treated in this simplified way, so
avoiding the dyadic method~\cite{Matute}.

Actually, the usual technique for solving boundary-value problems
introduces the electromagnetic potentials as intermediary field
quantities.  These are defined by~\cite{Jackson}
\begin{equation}
{\bf B} = {\bf \nabla} {\bf \times} {\bf A} , \hspace{0.5cm} {\bf
E} = - {\bf \nabla} \phi - \frac{\partial {\bf A}}{\partial t} ,
\end{equation}
with the subsidiary Lorentz condition
\begin{equation}
{\bf \nabla} \cdot {\bf A} + \epsilon \mu \;
\frac{\partial \phi}{\partial t} = 0 .
\end{equation}
It is then found that these potentials satisfy the inhomogeneous wave
equations
\begin{eqnarray}
& \nabla^{2} \phi - \epsilon \mu \; \displaystyle \frac{\partial^2
\phi}{\partial t^2} = - \frac{\rho}{\epsilon} , & \nonumber \\ & &
\nonumber \\ & \nabla^{2} {\bf A} - \epsilon \mu \; \displaystyle
\frac{\partial^2 {\bf A}}{\partial t^2} = - \mu {\bf J} , &
\end{eqnarray}
which together with the Lorentz condition form a set of equations
equivalent to the Maxwell equations.  The boundary conditions for
the potentials may be deduced from Eq.~(\ref{boundary}).

For fields that vary with an angular frequency $\omega$, i.e.
\begin{equation}
\phi({\bf r},t) = \phi({\bf r}) e^{-i \omega t} , \hspace{0.5cm}
{\bf A}({\bf r},t) = {\bf A}({\bf r}) e^{-i \omega t} ,
\end{equation}
we get equations that do not depend on time in regions free of charge
and current:
\begin{eqnarray}
& \nabla^{2} \phi + k^2 \; \phi = 0 , & \nonumber \\
& & \nonumber \\ & \nabla^{2} {\bf A} + k^2 \; {\bf A} = 0 , &
\label{pots}
\end{eqnarray}
which are like those in Eq.~(\ref{Helmholtz}) for the electric and
magnetic induction fields, so that in general we also confront,
for the vector potential, the mathematical complexities mentioned
above for the electromagnetic fields.

The purpose of this paper is to get general solutions of the
electromagnetic vector equations in spherical coordinates with
azimuthal symmetry using separation of variables in spite of
having equations that mix field components.  Boundary
conditions are easier to apply to these solutions, and their forms
highlight the similarities and differences between the electric
and magnetic cases in both time-independent and time-dependent
situations. The approach shows that boundary-value problems can be
solved for the electric and magnetic vector fields directly, and
that the process involves the same kind of mathematics as the
usual approach of solving for potentials.  The material in this
work may be used in a beginning graduate course in classical
electromagnetism or mathematical methods for physicists.  It is
organized as follows.  In Sec.~2 we describe the method for the
static case showing how the mathematical complications of solving
the vector field equations are easily overcome by means of
separation of variables.  In Sec.~3 we extend the method to
discuss the case of time-varying fields.  Concluding remarks are
given in Sec.~4.

\section{Static fields}
For steady-state electric and magnetic phenomena, the fields outside
sources satisfy the vector Laplace equations
\begin{equation}
\nabla^{2} {\bf E} = {\bf 0} , \hspace{0.5cm} \nabla^{2} {\bf B} =
{\bf 0} ,
\end{equation}
where only transverse components with zero divergence are
involved. Supposing all the charge and current are on the bounding
surfaces, solutions in different regions can be connected through
the boundary conditions indicated in Eq.~(\ref{boundary}).  To
demonstrate the features of the treatment, we first consider
boundary-value problems with azimuthal symmetry in electrostatics.
The solution of stationary current problems in magnetostatics is
mathematically identical.

Combining the expressions for ${\bf \nabla} {\bf \times} ({\bf
\nabla} {\bf \times} {\bf E}) = {\bf 0}$ and ${\bf \nabla} \cdot
{\bf E} = 0$ in spherical coordinates and assuming no
$\varphi$-dependence, we find using Eq.~(\ref{nablas}) that the
components of the electric field $E_{r}$ and $E_{\theta}$ satisfy
the equations
\begin{eqnarray}
(\nabla^{2} {\bf E})_{r} & = & \displaystyle \frac{1}{r^{2}} \;
\frac{\partial^{2}}{\partial r^{2}} (r^{2} E_{r}) \nonumber \\ & +
& \frac{1}{r^{2} \sin \, \theta} \; \frac{\partial}{\partial
\theta} (\sin \, \theta \; \frac{\partial E_{r}}{\partial \theta})
\, = \, 0 ,
\label{radial} \\ & & \nonumber \\
(\nabla^{2} {\bf E})_{\theta} & = & \displaystyle \frac{1}{r} \;
\frac{\partial^{2}}{\partial r^{2}} (r E_{\theta}) - \frac{1}{r}
\; \frac{\partial^2 E_{r}}{\partial r
\partial \theta} \, = \, 0 .
\label{angular}
\end{eqnarray}
Equation (\ref{radial}) is for $E_{r}$ alone, whereas
Eq.~(\ref{angular}) involves both components.  There is also a
separated equation for $E_{\varphi}$:
\begin{eqnarray}
(\nabla^{2} {\bf E})_{\varphi} & = & \frac{1}{r} \;
\frac{\partial^{2}}{\partial r^{2}} (r E_{\varphi}) +
\frac{1}{r^{2} \sin \, \theta} \; \frac{\partial}{\partial \theta}
\nonumber \\ & \times & (\sin \, \theta \; \frac{\partial
E_{\varphi}}{\partial \theta}) - \, \frac{1}{r^{2} \sin^{2}
\theta} \; E_{\varphi} \, = \, 0 .
\label{phiField}
\end{eqnarray}
In this paper, however, we will not be concerned about those
cylindrical symmetry cases where only the $\varphi$-component of
the vector field is nonzero because a scalar technique of
separation of variables is already known to obtain the
solution~\cite{Arfken}.

Using the transverse condition
\begin{equation}
{\bf \nabla} \cdot {\bf E} =
\frac{1}{r^{2}} \; \frac{\partial}{\partial r} (r^{2} E_{r}) +
\frac{1}{r \sin \, \theta} \; \frac{\partial}{\partial \theta}
(\sin \, \theta \; E_{\theta}) = 0 ,
\label{div}
\end{equation}
where azimuthal symmetry is assumed, Eq.~(\ref{radial}) implies
\begin{equation}
\frac{\partial}{\partial r} (r E_{\theta}) -
\frac{\partial E_{r}}{\partial \theta} = 0 ,
\label{good}
\end{equation}
which is consistent with Eq.~(\ref{angular}).  Thus, to obtain
$E_{\theta}$ from $E_{r}$ we can consider either Eq.~(\ref{div})
or Eq.~(\ref{good}).  These equations correspond to choosing a
gauge when this method is applied to the vector potential.

Now, in order to solve Eq.~(\ref{radial}) for $E_{r}$, we refer to
the method of separation of variables and write the product form
\begin{equation}
E_{r}(r,\theta) = \frac{u(r)}{r^2} \; P(\theta) ,
\end{equation}
which leads to the following separated differential equations:
\begin{eqnarray}
& \displaystyle \frac{d^{2}u}{dr^{2}} - \frac{n(n+1)}{r^{2}} \; u = 0 , &
\label{req} \\ & & \nonumber \\
& \displaystyle \frac{1}{\sin \, \theta} \; \frac{d}{d \theta}
(\sin \, \theta \; \frac{d P}{d \theta}) + n (n + 1) P = 0 , &
\label{Legendre}
\end{eqnarray}
where $n(n+1)$ is the separation constant.  The solution of
Eq.~(\ref{req}) is
\begin{equation}
u(r) = a \, r^{n+1} + \frac{b}{r^{n}} ,
\end{equation}
where $a$ and $b$ are arbitrary constants.  Equation (\ref{Legendre})
is the Legendre equation of order $n$ and the only solution which is
single valued, finite and continuous over the whole interval corresponds
to the Legendre polynomial $P_{n}(\cos \, \theta)$, $n$ being restricted
to positive integer values.  Thus the general solution for $E_{r}$ is
\begin{equation}
E_{r}(r,\theta) = \sum_{n=0}^{\infty} \left( a_{n} r^{n-1} +
\frac{b_{n}}{r^{n+2}} \right) P_{n}(\cos \, \theta) .
\label{Er}
\end{equation}
The simplest way of solving Eq.~(\ref{div}) for $E_{\theta}$ is to
use the series expansion
\begin{equation}
E_{\theta}(r,\theta) = \sum_{n=0}^{\infty} v_{n}(r) \; \frac{d}{d\theta}
P_{n}(\cos \, \theta) ,
\label{Etheta}
\end{equation}
where $v_{n}(r)$ are functions to be determined.  By replacing
Eqs.~(\ref{Er}) and (\ref{Etheta}) into Eq.~(\ref{div}), it is
found that
\begin{equation}
v_{n}(r) = \frac{a_{n}}{n} \; r^{n-1} - \frac{b_{n}}{n+1} \;
\frac{1}{r^{n+2}}
\label{v}
\end{equation}
for $n \geq 1$ with $a_{o} = 0$; this null factor in
Eq.~(\ref{Er}) means the absence of static field terms of the
$1/r$ type, which are in reality typical of radiative fields as
shown below.  Clearly, the solutions given in Eqs.~(\ref{Er}),
(\ref{Etheta}) and (\ref{v}) satisfy Eq.~(\ref{good}). The
coefficients $a_{n}$ and $b_{n}$ are to be determined from the
boundary conditions.  For completeness, we include here the
well-behaved general solution of Eq.~(\ref{phiField}):
\begin{equation}
E_{\varphi}(r,\theta) = \sum_{n=0}^{\infty} \left( c_{n} r^{n} +
\frac{d_{n}}{r^{n+1}} \right) \frac{d}{d\theta} P_{n}(\cos \, \theta) .
\label{Ephi}
\end{equation}
Thus, Eqs.~(\ref{Er})-(\ref{Ephi}) formally give all three components
of the electric field.  The same type of equations applies in
magnetostatics.  However, the boundary conditions of Eq.~(\ref{boundary})
will make the difference, implying in particular that $b_{\circ}=0$ in
the series expansion of Eq.~(\ref{Er}) in magnetostatics; this being
primarily related to the absence of magnetic monopoles.

To illustrate the use of the above formulas, we consider the
simple example of the electric field due to a ring of radius $a$
with total charge $Q$ uniformly distributed and lying in the
$x$-$y$ plane. It is usually solved through the scalar potential
method by using the result of the potential along the $z$-axis
obtained from Coulomb's law~\cite{Jackson+}.  The surface charge
density on $r=a$, localized at $\theta = \pi / 2$, is written as
\begin{equation}
\rho_{S}(a,\theta) = \frac{Q}{2 \pi a^2} \;
\delta(\cos \, \theta) ,
\label{source1}
\end{equation}
which may be expanded using the well-known Legendre series
\begin{equation}
\delta(\cos \, \theta) = \sum_{n=0}^{\infty} \; \frac{2n+1}{2} \;
P_{n}(0) \; P_{n}(\cos \, \theta) ,
\label{delta1}
\end{equation}
with $P_{n}(0)$ given by
\begin{equation}
P_{2n+1}(0) = 0 , \;\;\;
P_{2n}(0) = \frac{(-1)^n (2n+1)!}{2^{2n} (n!)^2} .
\end{equation}
Taking into account the cylindrical symmetry of the system, and
the requirement that the series solutions in
Eqs.~(\ref{Er})-(\ref{v}) have to be finite at the origin, vanish
at infinity and satisfy the boundary conditions of
Eq.~(\ref{boundary}) at $r=a$ for all values of the angle
$\theta$, namely, $E_{\theta}$ continuous at $r=a$ and $E_{r}$
discontinuous at $r=a$, it is straightforwardly found that the
spherical components of the electric field are
\begin{eqnarray}
E_{r}(r,\theta) & = & \frac{Q}{4 \pi \epsilon_{\circ} r^2}
\sum_{n=0}^{\infty} P_{n}(0) \; P_{n}(\cos \, \theta) \nonumber
\\ & & \times \left\{ \begin{array}{l}
        \displaystyle (n+1) \left( \frac{a}{r} \right)^{n}
        \; , \; r > a \\  \\
        \displaystyle  - n \left( \frac{r}{a} \right)^{n+1} \; , \; r < a
        \end{array}
\right.
\label{Er1}
\end{eqnarray}

\begin{eqnarray}
E_{\theta}(r,\theta) & = & - \frac{Q}{4 \pi \epsilon_{\circ} r^2}
\sum_{n=0}^{\infty} P_{n}(0) \; P_{n}^{1}(\cos \, \theta)
\nonumber \\ & & \nonumber \\ & & \times \left\{ \begin{array}{l}
        \displaystyle \left( \frac{a}{r} \right)^{n}
        \; , \; r > a \\  \\
        \displaystyle \left( \frac{r}{a} \right)^{n+1} \; , \; r < a
        \end{array}
\right.
\end{eqnarray}
and $E_{\varphi}=0$, where $P_{n}^{1}(\cos \, \theta) =
(d/d\theta) \, P_{n}(\cos \, \theta)$ is an associated Legendre
function. Note in particular that the coefficient $b_{\circ}$ in
Eq.~(\ref{Er}) becomes $Q / 4 \pi \epsilon_{\circ}$ for $r > a$,
as expected. Also, the discontinuity of the $n$th component of
$E_{r}$ in Eq.~(\ref{Er1}) at $r=a$ is connected according to
Eq.~(\ref{boundary}) with the corresponding component of the
surface charge density $\rho_{S}$ obtained from
Eqs.~(\ref{source1}) and (\ref{delta1}), exhibiting the unity of
the multipole expansions of fields and sources (see
Ref.~\cite{LeyKoo}).

To clarify the application of the formulas in the case of
magnetostatics and also compare with electrostatics, we consider
next the magnetic analog of the above example, that is, the
magnetic induction field from a circular current loop of radius
$a$ lying in the $x$-$y$ plane and carrying a constant current
$I$. The surface current density on $r=a$ can be written as
\begin{equation}
{\bf J}_{S}(a,\theta,\varphi) = \frac{I}{a} \; \delta(\cos \, \theta) \;
{\bf \hat{\varphi}} ,
\label{source2}
\end{equation}
where for the delta function is now convenient to use the expansion
\begin{equation}
\delta(\cos \, \theta) = \sum_{n=0}^{\infty} \;
\frac{2n+1}{2n(n+1)} \; P_{n}^{1}(0) \; P_{n}^{1}(\cos \, \theta)
, \label{delta2}
\end{equation}
which follows from the completeness relation for the spherical harmonics
after multiplication by $e^{-i\varphi}$ and integration over $\varphi$.
The values for $P_{n}^{1}(0)$ are
\begin{equation}
P_{2n}^{1}(0) = 0 , \;\;\;
P_{2n+1}^{1}(0) = \frac{(-1)^{n+1} (2n+1)!}{2^{2n} (n!)^2} .
\label{loop3}
\end{equation}
Because of the cylindrical symmetry of the system,
$B_{\varphi}=0$. By requiring that the field be finite at the
origin, vanish at infinity and satisfy the boundary conditions of
Eq.~(\ref{boundary}) at $r=a$, the series solutions in
Eqs.~(\ref{Er})-(\ref{v}) for the magnetic case lead to the
following radial and angular components of the magnetic induction
field:
\begin{eqnarray}
B_{r}(r,\theta) & = & - \frac{\mu_{\circ} I a^2}{2 r^3}
\sum_{n=0}^{\infty} P_{n}^{1}(0) \; P_{n}(\cos \, \theta)
\nonumber \\ & & \times
\left\{ \begin{array}{l}
        \displaystyle \left( \frac{a}{r} \right)^{n-1}
        \; , \; r > a \\  \\
        \displaystyle \left( \frac{r}{a} \right)^{n+2} \; , \; r < a
        \end{array}
\right.
\label{loop1}
\end{eqnarray}

\begin{eqnarray}
B_{\theta}(r,\theta) & = & \frac{\mu_{\circ} I a^2}{2 r^3}
\sum_{n=0}^{\infty} P_{n}^{1}(0) \; P_{n}^{1}(\cos \, \theta)
\nonumber \\ & & \nonumber \\ & & \times \left\{ \begin{array}{l}
        \displaystyle \frac{1}{n+1} \left( \frac{a}{r} \right)^{n-1}
        \; , \; r > a \\  \\
        \displaystyle - \frac{1}{n} \left( \frac{r}{a} \right)^{n+2} \; ,
        \; r < a
        \end{array}
\right.
\label{loop2}
\end{eqnarray}
Note that, as anticipated for magnetostatic problems, the
coefficient $b_{\circ}$ in Eq.~(\ref{Er}) is equal to zero. Also,
as expected, the discontinuity of the $n$th component of
$B_{\theta}$ in Eq.~(\ref{loop2}) at $r=a$ is connected according
to Eq.~(\ref{boundary}) with the corresponding component of the
surface current density $J_{S \varphi}$ obtained from
Eqs.~(\ref{source2}) and (\ref{delta2}).  Another characteristic
difference with the electrostatic analog is that the coefficient
$P_{n}^{1}(0)$ appears instead of $P_{n}(0)$.  This can be traced
to the fact that the inhomogeneous boundary condition, as given by
Eq.~(\ref{boundary}), is applied to the angular component of the
magnetic induction field in Eqs.~(\ref{Etheta})-(\ref{v}), as
opposed to the corresponding inhomogeneous boundary condition
acting on the radial component of the electric field in
Eq.~(\ref{Er}).  The fields in Eqs.~(\ref{loop1})-(\ref{loop2})
are usually obtained through the vector potential method by using
the expression of the magnetic induction field along the $z$-axis
calculated from the Biot and Savart law~\cite{Arfken}. An
alternative technique is mere integration of the vector
potential~\cite{Jackson++}. Our treatment has the advantage of
introducing a considerable simplification on the procedure of
applying the boundary conditions on the magnetic induction field
directly.

\section{Time-varying fields}
By using Eqs.~(\ref{Maxwell}), (\ref{linear}) and (\ref{harmonic})
it is seen that outside sources the fields are related by
\begin{equation}
{\bf E} = \frac{i \omega}{k^2} \; {\bf \nabla} {\bf \times} {\bf B} ,
\end{equation}
so that we only need to solve Eq.~(\ref{Helmholtz}) for ${\bf B}$.
Alternatively, we can solve for ${\bf E}$, and obtain ${\bf B}$
through the expression
\begin{equation}
{\bf B} = - \, \frac{i}{\omega} \; {\bf \nabla} {\bf \times} {\bf E} .
\end{equation}
In the following, we choose to deal with the Helmholtz equation
for the magnetic induction field.  The reason is to exhibit
similarities and differences with the static case treated in
Sec.~2.

In the case of spherical boundary surfaces with azimuthal
symmetry, the $B_{r}$ and $B_{\theta}$ components of the magnetic
induction satisfy the following equations:
\begin{eqnarray}
(\nabla^{2} {\bf B})_{r} + k^2 \; B_{r} & = & \displaystyle
\frac{1}{r^{2}} \; \frac{\partial^{2}}{\partial r^{2}} (r^{2}
B_{r}) + \frac{1}{r^{2} \sin \, \theta} \nonumber \\ & \times &
\frac{\partial}{\partial \theta} (\sin \, \theta \; \frac{\partial
B_{r}}{\partial \theta}) + k^2 \; B_{r} \, = \, 0 , \hspace{0.8cm}
\label{Bradial}
\end{eqnarray}
\begin{eqnarray}
(\nabla^{2} {\bf B})_{\theta} + k^2 \; B_{\theta} & = &
\displaystyle \frac{1}{r} \; \frac{\partial^{2}}{\partial r^{2}}
(r B_{\theta}) - \frac{1}{r} \; \frac{\partial^2 B_{r}}{\partial r
\partial \theta} \nonumber \\ & + & k^2 \; B_{\theta} \, = \, 0 .
\label{Bangular}
\end{eqnarray}
Similarly, for the $B_{\varphi}$ component we would have the
equation
\begin{eqnarray}
(\nabla^{2} {\bf B})_{\varphi} + k^2 \; B_{\varphi} =
\displaystyle \frac{1}{r} \; \frac{\partial^{2}}{\partial r^{2}}
(r B_{\varphi}) + \frac{1}{r^{2} \sin \, \theta}
\frac{\partial}{\partial \theta} \hspace{0.5cm} & & \nonumber \\
\times \; (\sin \, \theta \; \frac{\partial B_{\varphi}}{\partial
\theta}) - \, \frac{1}{r^{2} \sin^{2} \theta} \; B_{\varphi} + k^2
\; B_{\varphi} \, = \, 0 . & &
\end{eqnarray}
These are analogous to Eqs.~(\ref{radial}), (\ref{angular}) and
(\ref{phiField}) in connection with the vector Laplace equation.
In order to solve Eq.~(\ref{Bradial}) we let
\begin{equation}
B_{r}(r,\theta) = \frac{j(r)}{r} \; P(\theta) ,
\end{equation}
whereupon separation yields
\begin{equation}
\frac{d^{2}j}{dr^{2}} + \frac{2}{r} \; \frac{dj}{dr} + \left[
k^{2} - \frac{n(n+1)}{r^{2}} \right] j = 0 ,
\label{Bessel}
\end{equation}
and Eq.~(\ref{Legendre}), where the constant $n(n+1)$ is the
separation parameter. Equation (\ref{Bessel}) is the spherical
Bessel equation of order $n$ with variable $kr$.  Therefore, the
general solution for $B_{r}$ is
\begin{equation}
B_{r}(r,\theta) = \sum_{n=0}^{\infty} \left[ a_{n} \frac{j_{n}(kr)}{r} +
b_{n} \frac{n_{n}(kr)}{r} \right] P_{n}(\cos \, \theta) .
\label{Hradial}
\end{equation}
Depending on boundary conditions, the spherical Hankel functions
$h_{n}^{(1,2)}$ instead of the spherical Bessel functions $j_{n}$,
$n_{n}$ may be used.  For $B_{\theta}$ we again write
\begin{equation}
B_{\theta}(r,\theta) = \sum_{n=0}^{\infty} w_{n}(r) \; \frac{d}{d\theta}
P_{n}(\cos \, \theta) ,
\label{Hangular}
\end{equation}
and use ${\bf \nabla \cdot B}=0$ to obtain now
\begin{eqnarray}
w_{n} & = & \displaystyle \frac{a_{n}}{n(n+1) r} \; \frac{d\;}{dr}
[r \, j_{n}(kr)] \nonumber \\ & + & \displaystyle
\frac{b_{n}}{n(n+1) r} \; \frac{d\;}{dr} [r \, n_{n}(kr)] ,
\label{w}
\end{eqnarray}
for $n \ge 1$ with $a_{\circ} = b_{\circ} =0$.
The other coefficients $a_{n}$ and $b_{n}$ are determined so that the
boundary conditions for the vector field are exactly satisfied.
In the case of the $B_{\varphi}$ component, the general solution is
\begin{eqnarray}
B_{\varphi}(r,\theta) = \sum_{n=0}^{\infty} \left[ c_{n} j_{n}(kr)
+ d_{n} n_{n}(kr) \right] \frac{d}{d\theta} P_{n}(\cos \, \theta)
. \hspace{0.3cm} & &
\end{eqnarray}
The same type of equations applies for the electric field.

As an example, we shall consider the problem of the magnetic
induction field from a current $I=I_{\circ} e^{-i \omega t}$ in a
circular loop of radius $a$ lying in the $x$-$y$ plane.  It is the
time-varying version of the case solved in Sec.~2.  The surface
density current on $r=a$ is then
\begin{equation}
{\bf J}_{S}(a,\theta,\varphi,t) = \frac{I_{\circ}}{a} \;
\delta(\cos \, \theta) \; e^{-i \omega t} \; {\bf \hat{\varphi}}  ,
\label{source3}
\end{equation}
which can be expanded using Eq.~(\ref{delta2}).
The complete series solution of the Helmholtz equation for the
magnetic induction field, which is finite at the origin,
represents outgoing waves at infinity and satisfies the boundary
conditions of Eq.~(\ref{boundary}) at $r=a$, becomes
\begin{eqnarray}
B_{r}(r,\theta,t) & = & -i \frac{\mu_{\circ} I_{\circ} k a}{2 r}
e^{-i \omega t} \sum_{n=0}^{\infty} (2n+1) P_{n}^{1}(0)
\nonumber \\ & & \times P_{n}(\cos \, \theta) \left\{
\begin{array}{l}
        \displaystyle j_{n}(ka) \; h_{n}^{(1)}(kr)
        \\  \\
        \displaystyle j_{n}(kr) \; h_{n}^{(1)}(ka)
        \end{array}
\right.
\label{mif1}
\end{eqnarray}

\begin{eqnarray}
& B_{\theta}(r,\theta,t) = \displaystyle -i \frac{\mu_{\circ}
I_{\circ} k^{2} a}{2} e^{-i \omega t} \sum_{n=0}^{\infty}
\frac{2n+1}{n(n+1)} P_{n}^{1}(0) & \nonumber \\ & \times
P_{n}^{1}(\cos \, \theta) \left\{ \begin{array}{l}
        j_{n}(ka) \left[ h_{n-1}^{(1)}(kr) - \displaystyle
        \frac{n}{kr} \; h_{n}^{(1)}(kr) \right]
        \\  \\
        h_{n}^{(1)}(ka) \left[ j_{n-1}(kr) - \displaystyle
        \frac{n}{kr} \; j_{n}(kr) \right]
        \end{array}
\right. &
\label{mif2}
\end{eqnarray}
and $B_{\varphi}=0$, where the upper line holds for $r>a$ and the
lower line for $r<a$.  As noted above, the coefficient $a_{\circ}$
in Eq.~(\ref{Hradial}) indeed vanishes.  Also, the discontinuity
of the $n$th component of $B_{\theta}$ in Eq.~(\ref{mif2}) at
$r=a$ is connected, according to Eq.~(\ref{boundary}), with the
$n$th component of the surface current density $J_{S \varphi}$
obtained from Eqs.~(\ref{source3}) and (\ref{delta2}).  A
characteristic difference between this time-varying problem and
the corresponding static case is the appearance of the spherical
Bessel functions, which are solutions of the radial part of the
Helmholtz equation in spherical coordinates.  Using their limiting
values~\cite{Arfken+}, it can be seen that for $k \rightarrow 0$
the static results in Eqs.~(\ref{loop1}) and (\ref{loop2}) are
obtained, as mathematically and physically expected. On the other
hand, the radiative part of the external magnetic induction field,
which decreases as $1/r$, is given by
\begin{eqnarray}
{\bf B}(r,\theta,t) & = & {\bf \hat{\theta}} \; \frac{\mu_{\circ}
I_{\circ} k a}{4 r} e^{i(k r - \omega t)} \sum_{n=0}^{\infty}
\frac{(4n+3)(2n-1)!}{2^{2n}n!(n+1)!} \; \nonumber \\ & & \times
j_{2n+1}(ka) \; P_{2n+1}^{1}(\cos \, \theta) .
\end{eqnarray}
In the dipole approximation, $ka \ll 1$, this becomes the radiative
magnetic induction field from an oscillating magnetic dipole of magnetic
moment ${\bf m} = \pi a^2 I_{\circ} {\bf \hat{z}}$:
\begin{equation}
{\bf B}({\bf r},t) = \frac{\mu_{\circ} k^2}{4 \pi} \; ({\bf
\hat{r}} {\bf \times} {\bf m}) {\bf \times} {\bf \hat{r}} \;
\displaystyle \frac{e^{i (k r - \omega t)}}{r}  .
\end{equation}
The magnetic induction field in Eqs.~(\ref{mif1}) and (\ref{mif2})
can be seen to be just that which is obtained with the more
arduous technique of a dyadic Green's function expanded in vector
spherical harmonics and applied to the vector potential, which, by
symmetry, only has the $\varphi$-component different from
zero~\cite{Morse}. As we have shown, a direct calculation of the
electromagnetic field with $r$- and $\theta$-components is much
simplified if separation of variables is used.

\section{Conclusion}
For spherical coordinate systems, the Maxwell equations outside
sources lead to coupled equations involving all three components
of the electromagnetic fields.  In general, the statement is that
one cannot separate spherical components of the Maxwell equations,
and extensive techniques for solving the vector equations have
been developed which introduce vector spherical harmonics and use
dyadic methods.  We have shown, however, that separation of
variables is still possible in the case of azimuthal symmetry, and
so general solutions for each component of the electromagnetic
vector fields were obtained.  We have illustrated the use of these
formulas with direct calculations of electric and magnetic
induction fields from localized charge and current distributions,
without involving the electromagnetic potentials. Boundary
conditions are easier to apply to these solutions, and their forms
highlight the similarities and differences between the electric
and magnetic cases in both time-independent and time-dependent
situations.  Finally, we remark that in cylindrical coordinates,
the other commonly used curvilinear coordinate system, the Maxwell
equations do separate into equations for each vector component
alone if there is cylindrical symmetry, so that the method of
separation of variables can be used directly.

\begin{acknowledgments}
This work was partially supported by the Departamento de
Investigaciones Cient\'{\i}ficas y Tecnol\'ogicas, Universidad de
Santiago de Chile.
\end{acknowledgments}

\end{document}